\documentclass[prl,twocolumn,showpacs]{revtex4}
\usepackage{graphicx}
\usepackage{graphics}
\usepackage{amsfonts}
\usepackage{amsmath}
\usepackage{epsfig}

\begin{document}
\catcode`\ä = \active \catcode`\ö = \active \catcode`\ü = \active
\catcode`\Ä = \active \catcode`\Ö = \active \catcode`\Ü = \active
\catcode`\ß = \active \catcode`\é = \active \catcode`\è = \active
\catcode`\ë = \active \catcode`\ô = \active \catcode`\ê = \active
\catcode`\ø = \active \catcode`\ò = \active \catcode`\í = \active
\catcode`\Ó = \active \catcode`\ú = \active \catcode`\á = \active
\catcode`\ã = \active
\defä{\"a} \defö{\"o} \defü{\"u} \defÄ{\"A} \defÖ{\"O} \defÜ{\"U} \defß{\ss} \defé{\'{e}}
\defè{\`{e}} \defë{\"{e}} \defô{\^{o}} \defê{\^{e}} \defø{\o} \defò{\`{o}} \defí{\'{i}}
\defÓ{\'{O}} \defú{\'{u}} \defá{\'{a}} \defã{\~{a}}



\newcommand{\li}{$^6$Li}
\newcommand{\na}{$^{23}$Na}
\newcommand{\cs}{$^{133}$Cs}
\newcommand{\kk}{$^{40}$K}
\newcommand{\rb}{$^{87}$Rb}
\newcommand{\vect}[1]{\mathbf #1}
\newcommand{\g}{g^{(2)}}
\newcommand{\one}{|1\rangle}
\newcommand{\two}{|2\rangle}
\newcommand{\V}{V_{12}}
\newcommand{\kfa}{\frac{1}{k_F a}}

\title{Observation of Feshbach resonances between two different atomic species}

\author{C.A. Stan, M.W. Zwierlein, C.H. Schunck, S.M.F. Raupach, and W. Ketterle}

\affiliation{Department of Physics\mbox{,} MIT-Harvard Center for Ultracold Atoms\mbox{,} and Research Laboratory of Electronics,\\ MIT, Cambridge, MA 02139}

\date{\today}

\begin{abstract}
We have observed three Feshbach resonances in collisions between \li\ and \na\ atoms. The resonances were identified as narrow loss features when the magnetic field was varied. The molecular states causing these resonances have been identified, and additional \li-\na\ resonances are predicted. These resonances will allow the study of degenerate Bose-Fermi mixtures with adjustable interactions, and could be used to generate ultracold heteronuclear molecules.
\end{abstract}

\maketitle

Feshbach resonances \cite{inou98,cour98fesh,stwa76,ties93} have made it possible to control interactions in ultracold atomic gases. By tuning the magnetic field near a value where the energy of two free atoms coincides with a molecular bound state, the sign and strength of the atomic interactions can be varied.  Such tunable interactions were used to Bose-Einstein condense atomic species with unfavorable collisional properties \cite{corn00JILA_collapse,khay02soliton,webe03}, for cooling fermionic mixtures to degeneracy \cite{ohar02science}, and to produce bright solitons \cite{khay02soliton,stre02soliton}. Measurements of Feshbach resonances led to precise determinations of interatomic potentials \cite{leo00}. An important recent application of Feshbach resonances was the production of ultracold molecules from ultracold atoms \cite{rega03mol}, and Bose-Einstein condensation of molecules \cite{grei03mol_bec,joch03bec,bart04crossover,zwie03molBEC}.

So far, all experiments on Feshbach resonances studied collisions between two atoms of the same species. A few theoretical papers predicted Feshbach resonances between different atomic species \cite{burk98feshbach, simo03feshbach}, but these have not been observed.  Interspecies Feshbach resonances should lead to a host of new scientific phenomena, including the study of ultracold Fermi-Bose mixtures with tunable interactions, for which boson-mediated Cooper pairing \cite{heis00,bijl00}, phase separation \cite{molm98phase} and supersolid order \cite{buch03} have been predicted. These resonances may also be used to produce polar molecules, at phase-space densities higher than those obtained by heteronuclear photoassociation \cite{shaf99,schl01,manc04KRbground,kerm04RbCs}. Ultracold polar molecules could be used for quantum computation \cite{demi02} for studies of correlated many-body systems \cite{gora02,bara02}, and for searches for an electronic dipole moment \cite{kozl95}.

In this work, we studied collisions between fermionic \li\ and bosonic \na. In the absence of any theoretical prediction, it was not clear if there were any resonances in the accessible range of magnetic fields. Three s-wave Feshbach resonances were observed  and assigned.

An ultracold mixture of \li\ in the $\left|F, m_F\right> = \left|3/2,3/2\right>$ and \na\ in the $\left|F, m_F\right> = \left|2,2\right>$ hyperfine states was produced by forced microwave evaporation of \na\ in a magnetic trap as previously described in \cite{hadz03big_fermi}. The evaporation was stopped before reaching quantum degeneracy, and the mixture was transferred into a single focus 1064 nm optical dipole trap with a waist of 25 $\mu$m and a maximum power of 9 W. The \li\ and \na\ atoms were then transferred to the $\left|1/2,1/2\right>$ and $\left|1,1\right>$ hyperfine states, respectively, by simultaneous RF Landau-Zener sweeps. The sweeps were done by ramping the magnetic field from 9 to 10 G in 10 ms while keeping the RF frequencies constant at 249.8 MHz for \li\ and 1790 MHz for \na\, yielding a transfer efficiency close to 100 \%. This mixture of atoms in their lower hyperfine states was further cooled to quantum degeneracy by ramping down the laser power of the optical trap from 9 W to 220 mW in 1.4s.
\begin{figure}
    \begin{center}
    \includegraphics[width=3.3in]{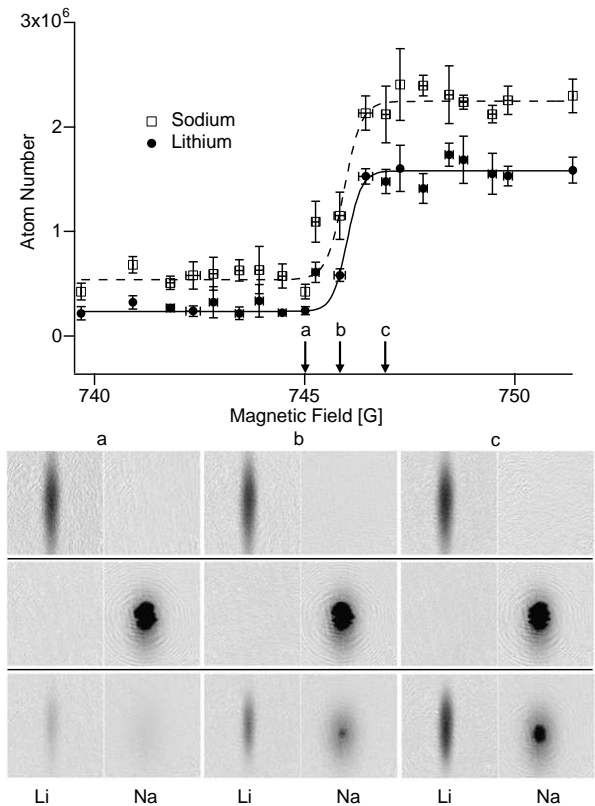}
    \caption[Title]{Determination of the position of the Feshbach resonance at 746 G. The magnetic field was swept down at a rate of 5.7 G/s from a field above the resonance to a variable final field. As the resonance was crossed, a sharp loss in atom number was observed. Top: \li\ (circles) and \na\ (open squares) atom numbers vs. the final field. The solid line (\li) and the dashed line (\na) are tangent hyperbolic fits to the data. Both fits locate the resonance at 746.0$\pm$0.4 G. A width of 0.8 G was obtained, which reflects both the intrinsic width of the resonances and the magnetic field stability. Bottom: The absorption images show the lithium cloud on the left side and the sodium cloud on the right.  They were taken at the three magnetic fields labeled a, b, c.  No losses were observed for lithium only (upper row) and sodium only (middle row), whereas the mixture (bottom row) showed the onset of losses of both species at the resonance. The sodium images clearly show the bimodal distribution characteristic for a Bose condensate. The field of view of each absorption image was 6 mm x 3 mm.} \label{fig:s-shapehigh}
    \end{center}
\end{figure}

Atom numbers were determined from absorption images of the atom clouds. For imaging, a second Landau-Zener sweep was used to transfer lithium and sodium back into their upper hyperfine states, where both species could be probed using cycling transitions. Lithium and sodium were imaged at the end of the experiment by taking first a lithium and then a sodium absorption image in rapid succession.  Typically, mixtures of $2\times 10^6$ degenerate fermions at $T/T_F \approx 0.25$ with a condensed \na\ cloud of $2\times 10^6$ atoms were produced at a temperature of 900 nK. The peak densities were $1 \times 10^{13} \rm\,cm^{-3}$ for \li\ and $2 \times 10^{14} \rm\,cm^{-3}$ for \na. The lifetime of the mixture was longer than 10 s.

We searched for Feshbach resonances up to the maximum magnetic field available experimentally, 1025 G, by sweeping the magnetic field across 20 G intervals at a rate of 21 G/s and determining the number of atoms left. The intervals containing resonances were identified by a higher loss rate. Precise measurements of the resonance positions were done by starting at a field near the resonance, and varying the endpoint of a constant rate sweep towards or across the resonance. The resonances were located by a sudden drop in the atom number, as seen in Fig.~\ref{fig:s-shapehigh} for the resonance at 746 G.  The magnetic field was calibrated by driving, at the resonance fields, RF transitions between the $\left|2\right>$ and  $\left|3\right>$, or between the $\left|2\right>$ and $\left|5\right>$, high field hyperfine states of \li.

In the same magnetic field range, we have observed one \li\ and six \na\ single species Feshbach resonances which have not been observed before \cite{schu04}. Interspecies \li -\na\ resonances were identified after checking that for a pure sample of either lithium or sodium, no resonant losses occurred at the same field. Three \li-\na\ resonances were observed, as listed in Table~\ref{tab:resonancepos}.

The losses are either due to (inelastic) three-body recombination, or due to resonant molecule formation during the sweep.  Other two-body loss mechanisms are not possible for \li\ and \na\ both in their lowest energy hyperfine state.  It should be noted that in the mixture near single species sodium Feshbach resonances, enhanced losses were also observed for lithium, presumably in a three-body process involving a sodium dimer and a lithium atom. Also, sodium losses were observed near the single species lithium resonance.

\begin{table}
\caption{\label{tab:resonancepos}Location of \li-\na\ Feshbach resonances and  the assignment of the molecular states which cause the resonances. Here $S$, $M_S$, $m_{i,\,\rm Li}$ and $m_{i,\,\rm Na}$ denote the total electronic spin, the magnetic quantum number for total electronic spin, and the magnetic quantum number for the \li\ and for the \na\ nuclear spin. Also listed are the ramp speeds for which 50\% of the atoms were lost. These values are a measure of the coupling strength between atomic and molecular levels.}
\begin{ruledtabular}
\begin{tabular}{ccc}
Magnetic Field&Ramp Speed&Molecular State\\
(G)&(G/s)&$\left|S\ M_S\ m_{\rm i,Li}\ m_{\rm i,Na}\right>$\\
\hline\\
$746.0\pm0.4$&15&$\left| 1\ \ \ 1\ \ -\rm 1\ \ \ 3/2\right>$\\
$759.6\pm0.2$&0.3&$\left| 1\ \ \ 1\ \ \ \ 0\ \ \ \ 1/2\right>$\\
$795.6\pm0.2$&10&$\left| 1\ \ \ 1\ \ \ \ 1\ -1/2\right>$\\
\end{tabular}
\end{ruledtabular}
\end{table}

\begin{figure*}
    \includegraphics[width=6in]{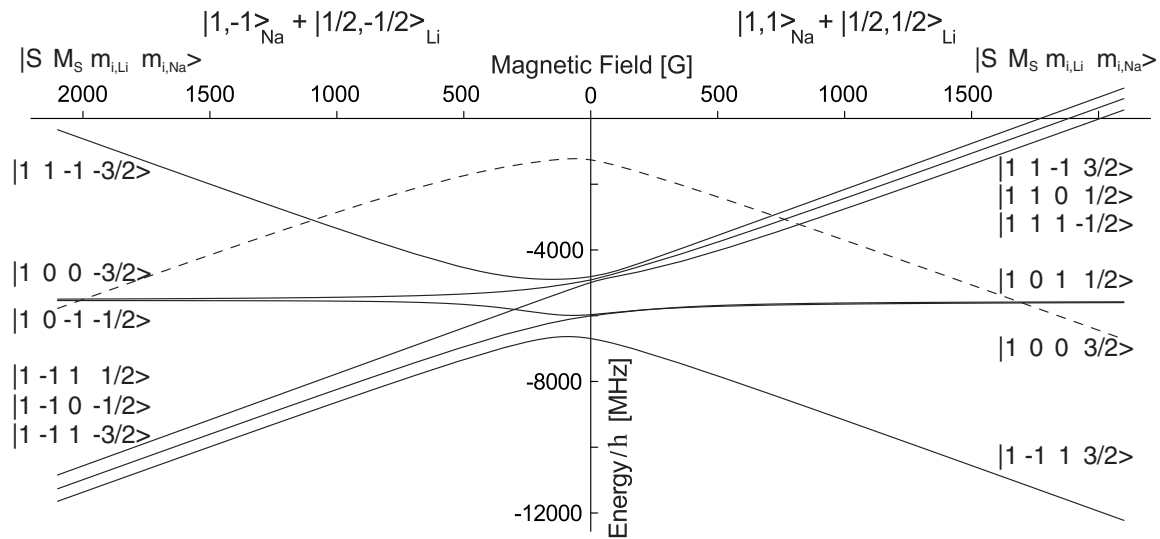}
    \caption{\label{LiNaResonance} Feshbach resonances in collisions between \li\ in $\left|1/2, \pm 1/2\right>$ and \na\ in $\left|1, \pm 1\right>$ hyperfine states. The solid lines show the energy dependence of triplet hyperfine molecular states for which coupling to the open channel is allowed by conservation of the total $M_F$. The dashed line shows the open channel energy threshold, equal to the sum of hyperfine energies of the incoming particle. The three observed Feshbach resonances occur in the $\left|1/2, 1/2\right>_{\rm Li} + \left|1, 1\right>_{\rm Na}$ collision channel, at the crossing of the threshold with three molecular states between 700 G and 850 G. For this channel, two other resonances should occur near 1700 G. In the $\left|1/2, - 1/2\right>_{\rm Li} + \left|1, - 1\right>_{\rm Na}$ collision channel, one resonance can be predicted to occur near 1100 G and another two near 2000 G.}
\end{figure*}

We assign the molecular states causing these resonances by extending the approach presented in~\cite{moer95res} to the case of two distinguishable atoms. Without the knowledge of the exact interatomic potentials the energies of the weakest bound molecular states cannot be predicted. However, one can still approximate the molecular hyperfine structure and Zeeman shifts and predict a pattern of Feshbach resonances with the molecular binding energy at zero magnetic field as the only adjustable parameter. We have checked that this procedure, applied to \na, reproduces all Feshbach resonances measured experimentally or predicted by coupled channels calculations to within 5\%.

The Coulomb interaction between two atoms naturally preserves their total electronic spin $\vec{S} = \vec{s}_1 + \vec{s}_2$. If we neglect the hyperfine interaction, $S$ is thus a good quantum number and the bound states are either singlet ($S=0$) or triplet ($S=1$). The molecular hyperfine energy at zero field is approximated as the sum of the hyperfine interaction for the two atoms:
$$H_{\rm hf} = h a_{\rm hf}^{\rm Li} \; \vec{s}_1 \cdot \vec{i}_{\rm Li} + h a_{\rm hf}^{\rm Na}\; \vec{s}_2 \cdot \vec{i}_{\rm Na},$$ with
$\vec{s}_{1,2}$ the spins of the two valence electrons, $\vec{i}_{\rm Li\, (Na)}$ the nuclear spin of the \li\ (\na) nucleus ($i_{\rm Li} = 1$, $i_{\rm Na} = 3/2$), and the hyperfine
 constants $a_{\rm hf}^{\rm Li} = 152.1368(1)\rm MHz$, $a_{\rm  hf}^{\rm Na} = 885.8131(1)\rm MHz$~\cite{arim77}. We rewrite $H_{\rm hf}$ in terms of $\vec{S}$:
\begin{eqnarray}
H_{\rm hf} &=&   \frac{h a_{\rm hf}^{\rm Li}}{2} \vec{S} \cdot \vec{i}_{\rm Li} + \frac{h a_{\rm hf}^{\rm Na}}{2} \vec{S} \cdot \vec{i}_{\rm Na}  \nonumber \\
 &&+ \frac{h a_{\rm hf}^{\rm Li}}{2} (\vec{s}_1-\vec{s}_2) \cdot \vec{i}_{\rm Li} + \frac{h a_{\rm hf}^{\rm Na}}{2}  (\vec{s}_1-\vec{s}_2) \cdot \vec{i}_{\rm Na} \nonumber \\
  & =&  V^+_{\rm hf} + V^-_{\rm hf}, \nonumber
\end{eqnarray}
Here, $V^+_{\rm hf}$ contains the terms proportional to $\vec{S}$, and thus preserves the separation of the orbital from the spin problem. $V^-_{\rm hf}$, however, mixes singlet and triplet states. Following~\cite{moer95res}, we will assume that the singlet-triplet spacing is large compared to $a_{\rm hf}^{\rm Na}$ and $a_{\rm hf}^{\rm Li}$, so that we can neglect $V^-_{\rm hf}$. In this approach, the orbital (Coulomb) part of the interaction remains decoupled from the spin part, which alone is responsible for the magnetic field dependence of the states.

For non-zero magnetic field, the spin part of the molecular hyperfine states is described by the Hamiltonian
$$ H^+_{\rm int} = V^+_{\rm hf} + V_{\rm Zeeman}$$
with
$$ V_{\rm  Zeeman} = \mu_B \vec{B} \cdot (g_S \vec{S} + g_{\rm Li} \,\vec{i}_{\rm Li} + g_{\rm Na}\, \vec{i}_{\rm Na}),$$
where $g_S$, $g_{\rm Li}$ and $g_{\rm Na}$ are the g-factors of the electron, of the \li, and of the \na\ nucleus. For collisions between $\left|1,1\right>_{\rm Na} + \left|1/2,1/2\right>_{\rm Li}$ the total $M_F = \frac{3}{2}$. In s-wave scattering, this incoming (open) channel can only be coupled to molecular states with the same $M_F$. In total, one can count two singlet and six triplet spin states with $M_F = \frac{3}{2}$.

Feshbach resonances occur at the magnetic fields at which a molecular hyperfine state has the same orbital plus hyperfine energy as the colliding atoms. The two almost degenerate singlet spin states cannot explain the existence of three well-separated resonances. Thus, the  observed resonances can be assigned to three triplet spin states with $M_S = 1$, differing only in their nuclear magnetic quantum numbers, as shown in Fig.~\ref{LiNaResonance}. Using the orbital energy $E_B$ as a free parameter, we performed a least-squares fit of the level crossings of these states with the atomic threshold, to the three experimentally obtained resonance positions. This resulted in $E_B = -5550 \pm 140$ MHz, which represents an estimate of the energy of a weakly bound triplet state of the \li-\na\ interatomic potential. The error was estimated by finding the values of $E_B$ which fit either the lowest ($746$G) or the highest ($796$G) resonance. The resulting resonant magnetic fields are 717 G, 763 G and 821 G. In view of the simplicity of the approach, the agreement with the observed resonance positions is satisfactory. Inverting the sign of the magnetic field, we obtain the level diagram for collisions of $\left|1,-1\right>_{\rm Na} + \left|1/2,-1/2\right>_{\rm Li}$ (Fig.~\ref{LiNaResonance}) which predicts several additional resonances.

The observed Feshbach resonances may be used to produce \li\na\ molecules by sweeping the magnetic field across the resonance. Atom pairs are converted into molecules by passing adiabatically through the Landau-Zener avoided crossing \cite{abee99,rega03mol}.  The efficiency of this conversion depends on the lifetime of the molecular cloud. In experiments with bosons, due to rapid vibrational relaxation, short molecular lifetimes on the order of a few ms have been observed. However, conversion efficiencies of a few \% could still be obtained \cite{durr04dissociation}.  The lifetime could be substantially increased by removing the leftover unbound atoms with a short light pulse \cite{xu03na_mol}.

Fermionic atoms offer the advantage of long molecular lifetimes due to Pauli suppression \cite{rega04lifetime,cubi03molecules,stre03}. In the case of \li$_2$ and \kk$_2$, highly vibrationally excited molecules were stable close to the resonance, where the molecules have a dominant atomic character \cite{petr03dimers}. Further away from the resonance, as the molecules become more strongly bound, the fermionic statistics of the constituents is no longer dominant, and loss rates increase \cite{cubi03molecules,rega04lifetime}.

We expect the opposite behavior for fermionic molecules produced near interspecies Feshbach resonances between a bosonic and fermionic atom.  Far away from the  resonance, all collisions between the molecules should be Pauli suppressed. Near the resonance, the collisions of long-range molecules are better approximated as collisions of their fermionic and bosonic atoms, and no significant suppression of collisions is expected. Therefore, a long-lived cloud of heteronuclear fermionic molecules could be realized by sweeping the magnetic field across and beyond the Feshbach resonance, and removing unbound atoms of both species with resonant light pulses.

In conclusion, we have observed three \li-\na\ Feshbach resonances and assigned them to weakly bound molecular states using a simple model. These resonances could be used for efficient production of ultracold heteronuclear molecules.

After this work was completed, the observation of Feshbach resonances in collisions between rubidium and potassium atoms was reported \cite{gold04damop}.

This work was supported by the NSF, ONR, ARO, and NASA. We thank Andrew Kerman for fruitful discussions and experimental assistance and Zoran Hadzibabic for a critical reading of the manuscript. S.\ Raupach is grateful to the Dr. J\"urgen Ulderup foundation for financial support.

\end{document}